\documentclass[aps,prl,notitlepage,reprint,superscriptaddress,nofootinbib,longbibliography]{revtex4-1}

\usepackage{graphicx}
\usepackage{dcolumn}
\usepackage{bm}
\usepackage{amsmath}
\usepackage{upgreek}
\usepackage[colorlinks,linkcolor=red,citecolor=blue,urlcolor=red]{hyperref}
\usepackage[utf8]{inputenc}
\usepackage[T1]{fontenc}
\usepackage{mathptmx}
\usepackage{ulem}
\usepackage[colorinlistoftodos]{todonotes}
\usepackage[mathlines]{lineno}% Enable numbering of text and display math
\renewcommand{\t}[1]{\mathrm{#1}}
\AtBeginDocument{%
	\newwrite\bibnotes
	\def\bibnotesext{Notes.bib}
	\immediate\openout\bibnotes=\jobname\bibnotesext
	\immediate\write\bibnotes{@CONTROL{REVTEX41Control}}
	\immediate\write\bibnotes{@CONTROL{%
			apsrev41Control,author="08",editor="1",pages="1",title="0",year="1"}}
	\if@filesw
	\immediate\write\@auxout{\string\citation{apsrev41Control}}%
	\fi
}%

\begin{document}
	
	\title{Ultrahigh-Q Torsional Nanomechanics through Bayesian Optimization}
	
	\author{A. D. Hyatt}
	\affiliation{Wyant College of Optical Sciences, University of Arizona, Tucson, AZ 85721, USA}
	
	\author{A. R. Agrawal}
	\affiliation{Wyant College of Optical Sciences, University of Arizona, Tucson, AZ 85721, USA}

        \author{C. M. Pluchar}
	\affiliation{Wyant College of Optical Sciences, University of Arizona, Tucson, AZ 85721, USA}

        \author{C. A. Condos}
	\affiliation{Wyant College of Optical Sciences, University of Arizona, Tucson, AZ 85721, USA}
 
	\author{D. J. Wilson}
	%\email{dalziel@optics.arizona.edu}
	\affiliation{Wyant College of Optical Sciences, University of Arizona, Tucson, AZ 85721, USA}
	
	\date{\today}
	\begin{abstract}
          \textcolor{black}{  
          Recently it was discovered that torsion modes of strained nanoribbons exhibit dissipation dilution, giving a route to enhanced torque sensing and quantum optomechanics experiments.  As with all strained nanomechanical resonators, an important limitation is bending loss due to mode curvature at the clamps.  Here we use Bayesian optimization to design nanoribbons with optimal dissipation dilution of the fundamental torsion mode.  Applied to centimeter-scale Si$_3$N$_4$ nanoribbons, we realize $Q$ factors exceeding $100$ million and $Q$-frequency products exceeding $ 10^{13}$ Hz at room temperature.  The thermal torque sensitivity of the reported devices is at the level of $10^{-20}\;\t{N}\,\t{m}/\sqrt{\t{Hz}}$ and the zero point angular displacement spectral density is at the level of $10^{-10}\;\t{rad}/\sqrt{\t{Hz}}$; they are moreover simple to fabricate, have high thermal conductivity, and can be heavily mass-loaded without diminishing their $Q$, making them attractive for diverse fundamental and applied weak force sensing tasks. }
	\end{abstract}
	% \vspace{-3mm}
	\maketitle

    Torsion oscillators have been instrumental to the advance of modern physics, from early investigations of gravity and the electrostatic force to the first observation of radiation pressure~\cite{Gillies1993}. Their low dissipation and resilience to environmental noise make them particularly attractive as weak force sensors \cite{adelberger2003tests,shaw2022torsion}; including more recently at the nanoscale \cite{kim2013nanoscale,pratt2023nanoscale}.  Applications of torsion nanoresonators include magnetometry \cite{davis2011nanomechanical}, radiometry \cite{cong2021chip}, and materials characterization \cite{kleiman1985single}.  A surge of interest in mass-loaded torsion microresonators is also now taking place, for short-range and quantum gravity tests \cite{manley2024microscale,PhysRevLett.116.221102,agafonova2024laser,bose2025massive}.

 \textcolor{black}{Recently it was shown \cite{pratt2023nanoscale} that strained nanoribbons exhibit torsion modes which are ``soft-clamped," \cite{tsaturyan2017ultracoherent} a form of modeshape engineering that enables ultrahigh quality $(Q)$ factors via dissipation dilution \cite{engelsen2024ultrahigh}.  Unlike previous demonstrations of soft-clamping, which rely on sophisticated phononic crystal \cite{tsaturyan2017ultracoherent,ghadimi2018elastic} and fractal suspensions \cite{bereyhi2022hierarchical}, in torsion modes it arises naturally due to the hierarchy of shear and bending stress. Besides simplicity, unique features of torsional dissipation dilution include the ability to soft-clamp the fundamental mode and to heavily mass-load it without diminishing its $Q$, both advantageous for sensing and transduction applications.}

    In this Letter we discuss the unique properties of torsion nanoribbons and how their dissipation dilution can be enhanced using Bayesian modeshape optimization.  Applied to centimeter-scale Si$_3$N$_4$ nanoribbons, we demonstrate, to our knowledge, the first solid state torsion oscillator with a $Q$ factor exceeding 100 million (``ultrahigh" $Q$).  The reported devices have $Q$-frequency products $>10^{13}$ Hz, meeting the condition for quantum coherence at room temperature.  They also have thermal torque sensitivities at the level of $10^{-20}\;\t{N}\,\t{m}/\sqrt{\t{Hz}}$.  We conclude with remarks on the future of optimized torsion nanoribbons for use in quantum optomechanics experiments and precision inertial sensing.

\textit{Torsional dissipation dilution and soft clamping} -- Strained flexural mode resonators exhibit enhanced $Q$ due to "dissipation dilution," an effect whereby the stiffness of an elastic body is increased without adding loss~\cite{engelsen2024ultrahigh}. Using a heuristic lumped mass model dating to Buckley (1914) \cite{buckley1914lxxxiv}, the fundamental torsion mode of a ribbon-like  beam (thickness $h$, width $w\gg h$, length $L\gg w$, elastic modulus $E$) under tension $T$ is predicted to experience a dissipation dilution factor \cite{pratt2023nanoscale}

\begin{equation}\label{eq:DQideal}
 \frac{Q}{Q_0}  = 1+\frac{k_\sigma}{k_E}\approx \frac{\sigma}{2E}\left(\frac{w}{h}\right)^2
\end{equation}
where $\sigma = T/(wh)$ is the tensile stress, %$E$ is the elastic modulus of the beam material, 
$k_\sigma = \sigma h w^3/(12 L)$ is the torsional rigidity of the beam due to tension \cite{buckley1914lxxxiv}, $k_E = Eh^3w/(6L)$ is the rigidity due to elastic stress (Saint-Venant's theorem \cite{barre1855memoire}), and $Q_0$ is the modal $Q$ in the absence of stress.

\begin{figure}[t]
%\vspace{-1mm}
\includegraphics[width=0.98\columnwidth]{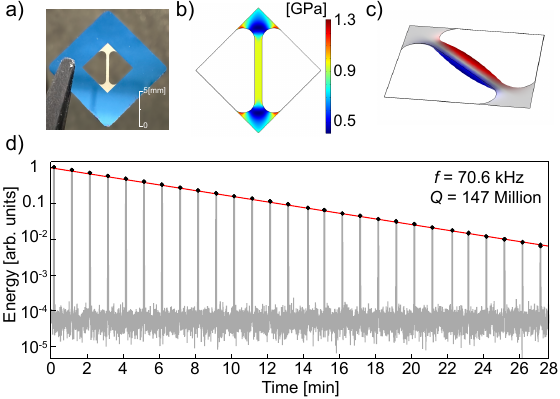}
		\caption{(a) Image of an optimized 7-mm-long, 400-$\mu$m-wide, 90-nm-thick Si$_3$N$_4$ nanoribbon. (b) Simulated Von Mises stress profile. (c) Simulated modeshape of fundamental torsion mode. (d) Stroboscopic energy ringdown of $f=70.6$ kHz torsion mode, yielding a quality ($Q$) factor of 150 million and a $Q$-$f$ product of $1.1\times10^{13}$ Hz.\label{fig:1}}
		\vspace{-9.5mm}
	\end{figure}

\begin{figure*}[ht!]
		\vspace{-2mm}
		\includegraphics[width=1.95\columnwidth]{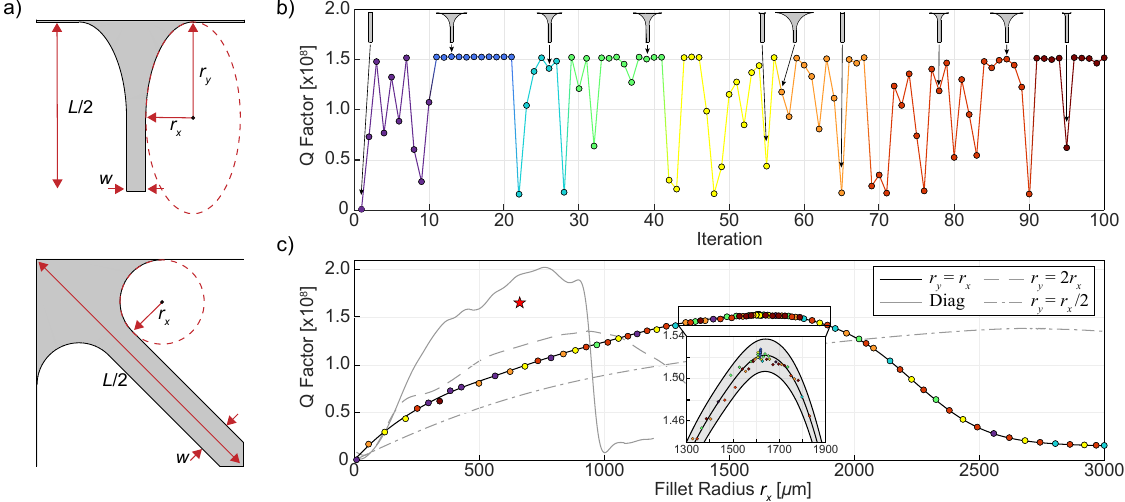}
		\caption{Optimizing torsional dissipation dilution by fillet design. (a) Two geometries are considered: elliptical fillet with perpendicular boundary (top), and circular fillet with diagonal boundary (bottom). Though an elliptical fillet is described by two parameters, we simplify to three special cases: $r_y = r_x$, $r_y = 2r_x$, and $r_y = r_x / 2$. (b) Demonstration of Bayesian algorithm's exploration vs exploitation behavior. After 10 random samples, the algorithm quickly converges on a possible solution. As it oversamples, an internal "counter" tells it to explore a new parameter regime. The plot in (b) is rearranged in (c) to reconstruct the objective function. Each colored point in (b) corresponds to (c). Several other traces of each geometry are plotted in gray with the hero sample from Fig \ref{fig:1} denoted by a red star for comparison.
\label{fig:2}}
		\vspace{-1mm}
	\end{figure*}
    
A modern continuum viscoelastic theory  \cite{fedorov2019generalized} reveals Eq.~\eqref{eq:DQideal} to be an ideal form of dissipation dilution type known as soft-clamping, in which the clamp region of the modeshape $u$ plays a negligible role in the elastic rigidity $k_E$. %\propto \int_0^w\int_0^L u_x''(x,y)dxdy$.  
For a uniform beam with fundamental torsion modeshape $u_\theta(z) \propto \sin(\pi z/L)-\pi \lambda (e^{-z/{\lambda L}}-1)$, a detailed calculation yields
\begin{equation}\label{eq:DQdetailed}
\frac{k_\sigma}{k_E}\approx   \frac{k_\sigma}{k_E^\t{shear,ext}+k_E^\t{bend,cl}}\approx \left(\frac{2E}{\sigma}\frac{h^2}{w^2}+\sqrt{\frac{E}{3\sigma}}\frac{h}{L}\right)^{-1},
\end{equation}
where $\lambda = (h/L)\sqrt{E/12\sigma}$, $k_E^\t{shear,ext}$ is the elastic rigidity due to shear strain in the extended region ($z\gg \lambda$) and $k_E^\t{bend,cl}$ is the rigidity due mode curvature near the clamps ($z\lesssim \lambda$). Equation \eqref{eq:DQdetailed} implies that clamping loss sets in when $w\gtrsim \sqrt{4\lambda}L$, yielding a maximum quality factor of
\begin{equation}\label{eq:QHC}
Q_\t{HC} = Q_0 \sqrt{\frac{3\sigma}{E}}\frac{L}{h} \approx 10^7 \sqrt{\frac{\sigma}{1\,\t{GPa}}\frac{250\,\t{GPa}}{E}}\frac{L}{3\,\t{mm}}\frac{Q_0/h}{60\,\t{nm}^{-1}}
\end{equation}
using typical stress, elastic modulus, and surface loss values for stoichiometric Si$_3$N$_4$ \cite{villanueva2014evidence}.  This ``hard-clamping'' limit is well-known in the study of tensioned nanomechanical beams and membranes (for which $Q_\t{HC} \rightarrow \sqrt{2}Q_\t{HC}$); however, it is seldom met due to extraneous forms of clamping loss.

\textcolor{black}{
\textit{Enhanced dissipation dilution due to clamp geometry -} 
Recently \cite{pratt2023nanoscale} we observed soft-clamping type dissipation dilution (Eq. \ref{eq:DQideal}) in the torsion modes of $L\approx 7\;\t{mm}$ , $w\approx 25-400\;\mu\t{m}$, $h\approx 75\;\t{nm}$ Si$_3$N$_4$ nanoribbons, similar to the device in Fig.~\ref{fig:1}, yielding $Q$ as high as $1.0\times 10^8$  for the widest ribbon---four times larger than the hard-clamping limit (Eq. \ref{eq:QHC}).  Finite element simulations reveal that this extended soft-clamping stems from the diagonal fillet geometry, which leads to a non-trivial modeshape and stress profile. Similar behavior has been observed and exploited in flexural modes of Si$_3$N$_4$ beams \cite{bereyhi2019clamp,sadeghi2019influence} and trampolines \cite{norte2016mechanical,reinhardt2016ultralow}. The key insight is that $k_E^\t{bend}$ depends on the integral of the mode curvature~\cite{tsaturyan2017ultracoherent, yu2012control}, 
 \begin{equation}\label{eq:kEbend_full}
 k_E^\t{bend} \propto \int (\nabla^2 u(x,y))^2 dxdy 
 \end{equation}
which depends sensitively on the beam cross-section and stress-profile near the clamps.
}

\textit{Fillet optimization via finite element simulation- } The anomalously high $Q$ observed in \cite{pratt2023nanoscale} motivates us to numerically search for a fillet geometry that minimizes Eq. \ref{eq:kEbend_full}, potentially extending torsional soft-clamping behavior ($Q>Q_\t{HC}$) into the ultrahigh-$Q$ regime $Q>10^8$.  
Towards this end, following an established approach \cite{ghadimi2018ultra}, we identify 
\begin{equation}\label{eq:DQnum}
 \frac{Q^{(n)}}{Q_0}  = 1+\frac{k_\sigma^{(n)}}{k_E^{(n)}}\approx \frac{K_\t{tot}^{(n)}}{U_\t{s}^{(n)}}
\end{equation}
where $K_\t{tot}^{(n)}$ and $U_\t{s}^{(n)}$ are the total kinetic energy and strain energy stored and in mode $n$, respectively \cite{sadeghi2019influence}.  We then determine $K_\t{tot}^{(n)}$ and $U_\t{s}^{(n)}$ from a two-step finite element model in COMSOL \cite{COMSOLuserguide}: a stationary study to find the stress redistribution and an eigenfrequency study to solve for the mode shapes and frequencies. ($K_\t{tot}^{(n)}$ and $U_\t{s}^{(n)}$ are stored in the model solution as "Wk \_tot" and "Ws\_tot", respectively \cite{COMSOLuserguide}.)

\begin{figure*}[ht!]
		\vspace{-1mm}
		\includegraphics[width=1.95\columnwidth]{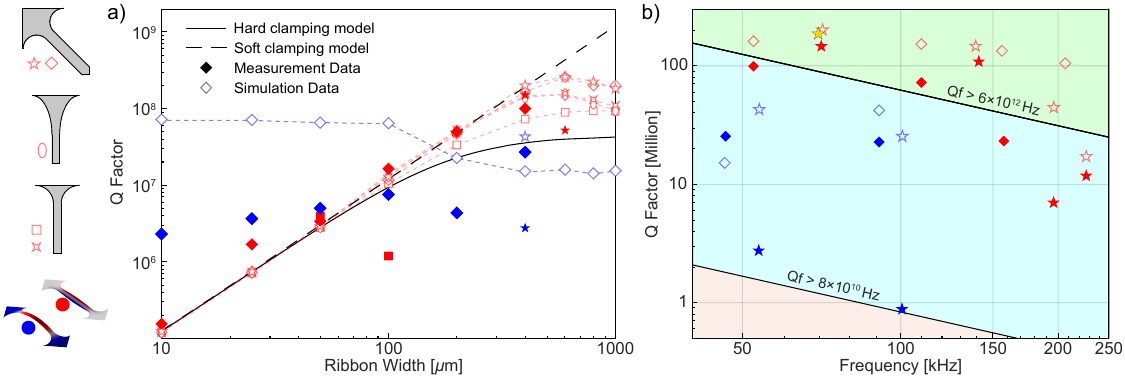}
		\caption{(a) Compilation of $Q$ for torsional (red) and flexural (blue) modes of $7$-mm-long, $90$-nm-thick ribbons of various widths $w$ and fillet geometries, including data from \cite{pratt2023nanoscale}. Open markers are finite-element simulations and solid markers are experimental data. Optimized ribbons are denoted by stars (diagonal geometry) and ellipses (perpendicular geometry).  Squares and diamonds correspond to non-optimized ribbons with fillet radius $r = w$. (b) Compilation of $Q\times f$ for $w = 400\,\mu$m ribbons with optimized and non-optimized fillets. Colored regions highlight the threshold for quantum coherence at 300 K (green) and 4 K (blue). The gold star corresponds to the measurement at 4 K shown in Fig. \ref{fig:4}. \label{fig:3}}
		\vspace{-1mm}
	\end{figure*}
             
    The above numerical technique predicts dissipation dilution factors $Q/Q_0$ for an arbitrary fillet geometry. Since it is non-analytical, however, we need to be creative to find an optimum geometry. We therefore turn to the field of inverse design, focusing on Bayesian optimization as described in Fig. \ref{fig:2}.
    
     \textit{Inverse design for dissipation engineering} Inverse design is a process in which one starts with desired performance requirements and computationally searches for a design that best fits those metrics. It poses the question of what design parameters provide the best performance. As such, the problem of inverse design is often approached using optimization algorithms to maximize or minimize some objective function (such as resonator $Q$), of which two stand out: topology optimization and Bayesian optimization. Topology optimization seeks to find the optimal mass distribution within a design space \cite{bendsoe2013topology} while Bayesian optimization finds the best set of parameters by efficiently sampling the search space \cite{mockus1991bayesian,snoek2012practical}.

    Both the topology \cite{hoj2021ultra,norder2024pentagonal} and the Bayesian approach \cite{shin2022spiderweb,cupertino2024centimeter} have a history in the field of nanomechanics. While topology optimization is more general, we use the Bayesian optimization algorithm because of its relative simplicity and availability. Originally developed in machine learning to optimize the hyperparameters of neural networks, the Bayesian optimization scheme excels at efficiently locating an approximate optimum of a black-box objective where the cost of evaluation is high \cite{pardalos2021black}. Its algorithm is flexible enough to work with both discrete and continuous variables while also avoiding local optima that masquerade as a global optimum value.

 Specifically, Bayesian optimization extremizes a black-box function by determining the best sampling points and estimating an optimum using a Gaussian process regression (GPR) \cite{wang2023intuitive}. The algorithm first takes a number of random samples to build an initial GPR model of the function. It then constructs and optimizes a simpler ``acquisition function" to determine the best point to sample next, based on a quantification of anticipated improvement \cite{gan2021acquisition} (accounting for the model confidence interval, current estimated optimum, or other fit parameters \cite{hoffman2011portfolio}). In our study, we use the ``expected improvement" acquisition function, which tracks both the probability that and by how much a set of parameters leads to an improved optimum \cite{jones1998expensive}.  The advantages of this approach are twofold: not only does the algorithm converge faster, but it is also able to escape local optima. The optimizer can be made further immune to local optima by including an "exploration ratio" term which regulates how much the algorithm over-samples certain areas of the search (a.k.a. ``exploration'') space \cite{gan2021acquisition}. 
 
\textit{Numerical results -} To use Bayesian optimization for inverse design, we simply call a COMSOL file evaluation as the objective function, feeding it different parameters and using Eq. \eqref{eq:DQnum} to find the $Q$ of the first-order torsional mode. For simplicity, we use MATLAB's built-in Bayesian optimization function with the "expected-improvement-plus" acquisition function type and tune the exploration ratio to 0.3. 

A numerical study of Bayesian optimized torsion nanoribbons is shown in Fig. \ref{fig:2}, for a nominal ribbon width, length, and thickness of $400\;\mu\t{m}$, $7\;\t{mm}$, and $90\;\t{nm}$, respectively, and material properties as in Eq. \ref{eq:QHC}.  We focus on the two fillet geometries shown in panel (a)---a ``perpendicular'' geometry where the fillet is terminated into the side a square window, and ``diagonal'' geometry in which the fillet is terminated into a corner of the square window---parameterized for simplicity by a single value \cite{MultipleParameter}. Figures \ref{fig:2}(b,c) highlight different regions of exploration versus exploitation by assigning each simulation (point) a color and grouping them to reveal the sampling patterns.  For both fillet geometries, the optimal quality factor $Q\approx 2\times 10^8$ exceeds the nominal hard-clamping limit $Q_\t{HC}$ by approximately an order of magnitude.

\textit{Experimental results - }To test the numerical simulations, we fabricated and characterized five 400-$\mu$m-wide and two 600-$\mu$m-wide ribbons with Bayesian-optimized diagonal fillets, an example of which is shown in Fig. \ref{fig:1}(a). The fabrication method is identical to that reported in \cite{pratt2023nanoscale}, and starts with a 200-$\mu$m-thick Si wafer coated with on both sides with Si$_3$N$_4$, followed by photolithography. After exposing the photomask, devices are released in KOH, cleaned by gradual dilution to methanol \cite{ghadimi2018ultra,agrawal2025ultra}, and dried (after removal from methanol) using a gentle breeze of compressed nitrogen~gas.

Quality factors were inferred from stroboscopic ringdown measurements as shown in Fig. \ref{fig:1}(d).  For these measurements, an optical lever \cite{pluchar2025quantum} was used as a displacement sensor and devices were housed in a high-vacuum chamber operating below 10$^{-7}$ mbar, to reduce gas damping~\cite{blom1992dependence}.   (Ringdowns were initiated by lightly tapping the chamber with a screwdriver.)  To reduce photothermal damping, the optical lever is only turned on for short intervals \cite{ghadimi2018elastic}, with a duty cycle of $<10\%$. 

A compilation of measured and simulated $Q$ factors versus ribbon width $w$ is shown in Fig. \ref{fig:3}, overlaid with the analytical model given in Eq. \ref{eq:DQdetailed} and experimental results from \cite{pratt2023nanoscale}. As is evident from the simulations, the analytical model is valid for sufficiently small widths that $Q\lesssim Q_\t{HC}$ (i.e. $w\lesssim \sqrt{4\lambda} L$).  For larger widths, the optimal $Q$ depends strongly on the fillet geometry, favoring diagonal fillets and saturating to an upper bound of $Q\approx 2.5\times 10^8$ for $w \approx 600\;\mu\t{m}$.

Of the six ribbons fabricated and characterized, we found that two devices---both 400-$\mu$m-wide---exhibited fundamental torsion modes with ultrahigh $Q$ factors: $Q  = 1.5\times 10^8$ and $1.2\times 10^8$, respectively, within a factor of two of the simulation in Fig. \ref{fig:2}(c). We studied the first device's higher-order modes, and found that both the fundamental and second order modes (with frequencies $f =70.6$ kHz and 139 kHz) exhibited $Q\times f > 10^{13}$~Hz, satisfying the condition for quantum coherence at room temperature, $Q\times f > k_B T/h_\t{P} = 6\times 10^\t{12}$ Hz, where $T = 293$ K, $k_\t{B}$ is Boltzmann's constant, and $h_\t{P}$ is Planck's constant.  We also studied the fundamental and second order flexural modes of the first device, and found them to be $\sim10$ times lower than the torsion mode and $\sim100$ times lower than simulated.  Indeed, for data reproduced from \cite{pratt2023nanoscale}, flexural mode $Q$ factors are consistently smaller than predicted, suggesting sensitivity to other forms of loss. 

\textit{Cryogenic torsional nanomechanics - }The high $Q$-$f$ product of our optimized torsion nanoribbons implies access to room temperature quantum experiments \cite{norte2016mechanical,wilson2009cavity}. It is also well-known that Si$_3$N$_4$ nanomechanical resonators exhibit increased $Q$ at cryogenic temperatures (especially below 100 mK \cite{gisler2022soft,yuan2015silicon,fischer2016optical}), leading to additional performance enhancement. Thus, a natural next step is to explore how optimized torsion nanoribbons behave in a cryogenic environment.

\begin{figure}[b!]
%		\vspace{-2mm}
		\includegraphics[width=1\columnwidth]{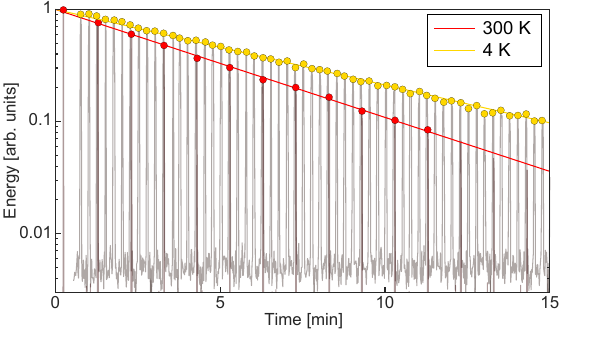}
		\caption{Torsion mode ringdowns of an optimized nanoribbon (identical to the device in Fig. \ref{fig:1}) at room temperature and $T\approx 4$ K, yielding $Q=1.2\times 10^8$ and $Q = 1.7\times 10^8$, respectively, corresponding a quantum coherence of $Qf/(k_B T/\hbar)\approx 1.4$ and $1.4\times 10^2$. \label{fig:4}}
		\vspace{-2mm}
	\end{figure}

We placed the second ultrahigh-$Q$ nanoribbon (identical to the device in Fig. \ref{fig:1}) in a high vacuum 4 K cryostat (Janis Systems ST-500-UHV), and conducted a ringdown of its fundamental torsion mode, as shown in Fig. \ref{fig:4}.  We observed a 40\% increase in its $Q$ relative to room temperature, from $1.2\times 10^8$ to $1.7\times10^8$, corresponding to a 100-fold increase in quantum coherence (the number of mechanical periods in the thermal decoherence time), from $Qf/(k_B T/h_\t{P})\approx 1.4$ to 140.

\textit{Summary $\&$ outlook - } In summary, we have used Bayesian optimization to design cm-scale Si$_3$N$_4$ nanoribbons with minimal torsion mode bending loss, enabling dissipation dilution factors exceeding the ``hard-clamping'' limit (Eq. \ref{eq:QHC}) by an order of magnitude, quality factors as high as $Q = 1.5\times 10^8$ ($1.7\times 10^8$) at room temperature (4 K) and $Q\times f>10^{13}$ Hz.

The extended soft-clamping and mesoscopic dimensions of the optimized nanoribbons make them attractive for weak torque sensing and quantum experiments, as can be seen from scaling laws for the resonance frequency $f \approx (2L)^{-1}\sqrt{\sigma/\rho}$, $Q$-$f$ product $Q f\approx Q_0w^2\sqrt{\sigma^3/\rho}/(4E Lh^2)$, moment of inertia $I\approx \rho L h w^3/24$, thermal torque power spectral density $S_\tau^\t{th} = 8\pi^2 k_B T I f/Q\propto h^3 w/Q_0$ and peak zero-point angular displacement spectral density $S^\t{ZP}_\theta =  \hbar  Q/(2I \pi^2 f^2)\propto Q_0/(h^3 w L)$  of the fundamental torsion mode, using Eq. \ref{eq:DQideal}.  For the device in Figs. \ref{fig:1} and \ref{fig:4}, for example, $I \approx 5\times 10^{-18} \;\t{kg}\,\t{m}^2$, $S_\tau^\t{th}(T =300\,\t{K})$ $\approx (1\times 10^{-20}\,\t{N m}/\sqrt{\t{Hz}})^2$, %$S_\tau^\t{th}(T=4\,\t{K})\approx (2\times 10^{-21}\,\t{N m}/\sqrt{\t{Hz}})^2$ 
and $S_\theta^\t{ZP}\approx (2\times 10^{-10}\,\t{rad}/\sqrt{\t{Hz}})^2$.  Combined with recent demonstrations of quantum-limited optical lever measurements with imprecisions of $S_\theta^\t{imp}\approx (10^{-11}\,\t{rad}/\sqrt{\t{Hz}})^2$ \cite{pluchar2025quantum,shin2024laser,hao2024back}, 
 for example, it should be possible to 
 feedback cool the torsion oscillator to a mean phonon number of $n_\t{m} \approx (2 n_\t{th} S_\theta^\t{imp}/S_\theta^\t{ZP})^{1/2}\sim 1$ from $T = 100$ mK, where $n_\t{th} = k_\t{B}T/(h_\t{P}f)$ is the thermal bath occupation~\cite{wilson2015measurement}.  Correspondingly, such a measurement would produce a quantum backaction torque of $S_\tau \ge \hbar^2/S_\theta^\t{imp}\sim (1\times 10^{-23}\,\t{N m}/\sqrt{\t{Hz}})^2$, within access of $S_\tau^\t{th}(T= 100\,\t{mK})$ \cite{hao2024back}.

 Finally, an attractive feature of strained nanoribbons is their ability to be loaded with a central pad without reducing the torsional $Q$ \cite{pratt2023nanoscale}. Combining a high-$Q$ nanosuspension with a tailored (materially and dimensionally) micro- to milligram scale torsion pad enables diverse sensing applications from magnetometry \cite{davis2011nanomechanical} and gravimetry \cite{condos2024ultralow} to, in its extreme form, searches for new physics such as Yukawa forces \cite{PhysRevLett.116.221102,manley2024microscale}, dark matter \cite{manley2024microscale}, and quantum gravity \cite{kryhin2023distinguishable,agafonova2024laser}.  Design of such detectors poses a challenging multi-parameter optimization problem, for which the Bayesian approach is well-suited.\\

\noindent\textit{Acknowledgments} - The authors thank Jon Pratt for insightful discussions and Mitul Dey Chowdhury,  Andrew Land, and Tyler ten Broek for assistance characterizing samples. This work was supported by National Science Foundation award no. 2239735. AH, ARA, CMP, and CAC acknowlege additional support from a Friends of Tucson Optics scholarship, a CNRS-UArizona iGlobes fellowship, the ARCS Foundation, and a UArizona National Labs Partnership Grant, respectively.

        \sloppy
        \bibliography{ref}

%merlin.mbs apsrev4-1.bst 2010-07-25 4.21a (PWD, AO, DPC) hacked
%Control: key (0)
%Control: author (8) initials jnrlst
%Control: editor formatted (1) identically to author
%Control: production of article title (0) allowed
%Control: page (1) range
%Control: year (1) truncated
%Control: production of eprint (0) enabled
\begin{thebibliography}{52}%
\makeatletter
\providecommand \@ifxundefined [1]{%
 \@ifx{#1\undefined}
}%
\providecommand \@ifnum [1]{%
 \ifnum #1\expandafter \@firstoftwo
 \else \expandafter \@secondoftwo
 \fi
}%
\providecommand \@ifx [1]{%
 \ifx #1\expandafter \@firstoftwo
 \else \expandafter \@secondoftwo
 \fi
}%
\providecommand \natexlab [1]{#1}%
\providecommand \enquote  [1]{``#1''}%
\providecommand \bibnamefont  [1]{#1}%
\providecommand \bibfnamefont [1]{#1}%
\providecommand \citenamefont [1]{#1}%
\providecommand \href@noop [0]{\@secondoftwo}%
\providecommand \href [0]{\begingroup \@sanitize@url \@href}%
\providecommand \@href[1]{\@@startlink{#1}\@@href}%
\providecommand \@@href[1]{\endgroup#1\@@endlink}%
\providecommand \@sanitize@url [0]{\catcode `\\12\catcode `\$12\catcode `\&12\catcode `\#12\catcode `\^12\catcode `\_12\catcode `\%12\relax}%
\providecommand \@@startlink[1]{}%
\providecommand \@@endlink[0]{}%
\providecommand \url  [0]{\begingroup\@sanitize@url \@url }%
\providecommand \@url [1]{\endgroup\@href {#1}{\urlprefix }}%
\providecommand \urlprefix  [0]{URL }%
\providecommand \Eprint [0]{\href }%
\providecommand \doibase [0]{http://dx.doi.org/}%
\providecommand \selectlanguage [0]{\@gobble}%
\providecommand \bibinfo  [0]{\@secondoftwo}%
\providecommand \bibfield  [0]{\@secondoftwo}%
\providecommand \translation [1]{[#1]}%
\providecommand \BibitemOpen [0]{}%
\providecommand \bibitemStop [0]{}%
\providecommand \bibitemNoStop [0]{.\EOS\space}%
\providecommand \EOS [0]{\spacefactor3000\relax}%
\providecommand \BibitemShut  [1]{\csname bibitem#1\endcsname}%
\let\auto@bib@innerbib\@empty
%</preamble>
\bibitem [{\citenamefont {Gillies}\ and\ \citenamefont {Ritter}(1993)}]{Gillies1993}%
  \BibitemOpen
  \bibfield  {author} {\bibinfo {author} {\bibfnamefont {G.~T.}\ \bibnamefont {Gillies}}\ and\ \bibinfo {author} {\bibfnamefont {R.~C.}\ \bibnamefont {Ritter}},\ }\bibfield  {title} {\enquote {\bibinfo {title} {Torsion balances, torsion pendulums, and related devices},}\ }\href {\doibase 10.1063/1.1144248} {\bibfield  {journal} {\bibinfo  {journal} {Rev. Sci. Instrum.}\ }\textbf {\bibinfo {volume} {64}},\ \bibinfo {pages} {283--309} (\bibinfo {year} {1993})},\ \Eprint {http://arxiv.org/abs/https://doi.org/10.1063/1.1144248} {https://doi.org/10.1063/1.1144248} \BibitemShut {NoStop}%
\bibitem [{\citenamefont {Adelberger}\ \emph {et~al.}(2003)\citenamefont {Adelberger}, \citenamefont {Heckel},\ and\ \citenamefont {Nelson}}]{adelberger2003tests}%
  \BibitemOpen
  \bibfield  {author} {\bibinfo {author} {\bibfnamefont {E.}~\bibnamefont {Adelberger}}, \bibinfo {author} {\bibfnamefont {B.}~\bibnamefont {Heckel}}, \ and\ \bibinfo {author} {\bibfnamefont {A.}~\bibnamefont {Nelson}},\ }\bibfield  {title} {\enquote {\bibinfo {title} {Tests of the gravitational inverse-square law},}\ }\href {https://www.annualreviews.org/content/journals/10.1146/annurev.nucl.53.041002.110503} {\bibfield  {journal} {\bibinfo  {journal} {Annual Review of Nuclear and Particle Science}\ }\textbf {\bibinfo {volume} {53}},\ \bibinfo {pages} {77--121} (\bibinfo {year} {2003})}\BibitemShut {NoStop}%
\bibitem [{\citenamefont {Shaw}\ \emph {et~al.}(2022)\citenamefont {Shaw}, \citenamefont {Ross}, \citenamefont {Hagedorn}, \citenamefont {Adelberger},\ and\ \citenamefont {Gundlach}}]{shaw2022torsion}%
  \BibitemOpen
  \bibfield  {author} {\bibinfo {author} {\bibfnamefont {E.}~\bibnamefont {Shaw}}, \bibinfo {author} {\bibfnamefont {M.}~\bibnamefont {Ross}}, \bibinfo {author} {\bibfnamefont {C.}~\bibnamefont {Hagedorn}}, \bibinfo {author} {\bibfnamefont {E.}~\bibnamefont {Adelberger}}, \ and\ \bibinfo {author} {\bibfnamefont {J.}~\bibnamefont {Gundlach}},\ }\bibfield  {title} {\enquote {\bibinfo {title} {Torsion-balance search for ultralow-mass bosonic dark matter},}\ }\href {https://doi.org/10.1103/PhysRevD.105.042007} {\bibfield  {journal} {\bibinfo  {journal} {Physical Review D}\ }\textbf {\bibinfo {volume} {105}},\ \bibinfo {pages} {042007} (\bibinfo {year} {2022})}\BibitemShut {NoStop}%
\bibitem [{\citenamefont {Kim}\ \emph {et~al.}(2013)\citenamefont {Kim}, \citenamefont {Doolin}, \citenamefont {Hauer}, \citenamefont {MacDonald}, \citenamefont {Freeman}, \citenamefont {Barclay},\ and\ \citenamefont {Davis}}]{kim2013nanoscale}%
  \BibitemOpen
  \bibfield  {author} {\bibinfo {author} {\bibfnamefont {P.~H.}\ \bibnamefont {Kim}}, \bibinfo {author} {\bibfnamefont {C.}~\bibnamefont {Doolin}}, \bibinfo {author} {\bibfnamefont {B.~D.}\ \bibnamefont {Hauer}}, \bibinfo {author} {\bibfnamefont {A.~J.}\ \bibnamefont {MacDonald}}, \bibinfo {author} {\bibfnamefont {M.~R.}\ \bibnamefont {Freeman}}, \bibinfo {author} {\bibfnamefont {P.~E.}\ \bibnamefont {Barclay}}, \ and\ \bibinfo {author} {\bibfnamefont {J.~P.}\ \bibnamefont {Davis}},\ }\bibfield  {title} {\enquote {\bibinfo {title} {Nanoscale torsional optomechanics},}\ }\href {https://pubs.aip.org/aip/apl/article/102/5/053102/1068573} {\bibfield  {journal} {\bibinfo  {journal} {Applied Physics Letters}\ }\textbf {\bibinfo {volume} {102}} (\bibinfo {year} {2013})}\BibitemShut {NoStop}%
\bibitem [{\citenamefont {Pratt}\ \emph {et~al.}(2023)\citenamefont {Pratt}, \citenamefont {Agrawal}, \citenamefont {Condos}, \citenamefont {Pluchar}, \citenamefont {Schlamminger},\ and\ \citenamefont {Wilson}}]{pratt2023nanoscale}%
  \BibitemOpen
  \bibfield  {author} {\bibinfo {author} {\bibfnamefont {J.~R.}\ \bibnamefont {Pratt}}, \bibinfo {author} {\bibfnamefont {A.~R.}\ \bibnamefont {Agrawal}}, \bibinfo {author} {\bibfnamefont {C.~A.}\ \bibnamefont {Condos}}, \bibinfo {author} {\bibfnamefont {C.~M.}\ \bibnamefont {Pluchar}}, \bibinfo {author} {\bibfnamefont {S.}~\bibnamefont {Schlamminger}}, \ and\ \bibinfo {author} {\bibfnamefont {D.~J.}\ \bibnamefont {Wilson}},\ }\bibfield  {title} {\enquote {\bibinfo {title} {Nanoscale torsional dissipation dilution for quantum experiments and precision measurement},}\ }\href {https://journals.aps.org/prx/abstract/10.1103/PhysRevX.13.011018} {\bibfield  {journal} {\bibinfo  {journal} {Phys. Rev. X}\ }\textbf {\bibinfo {volume} {13}},\ \bibinfo {pages} {011018} (\bibinfo {year} {2023})}\BibitemShut {NoStop}%
\bibitem [{\citenamefont {Davis}\ \emph {et~al.}(2011)\citenamefont {Davis}, \citenamefont {Vick}, \citenamefont {Li}, \citenamefont {Portillo}, \citenamefont {Fraser}, \citenamefont {Burgess}, \citenamefont {Fortin}, \citenamefont {Hiebert},\ and\ \citenamefont {Freeman}}]{davis2011nanomechanical}%
  \BibitemOpen
  \bibfield  {author} {\bibinfo {author} {\bibfnamefont {J.}~\bibnamefont {Davis}}, \bibinfo {author} {\bibfnamefont {D.}~\bibnamefont {Vick}}, \bibinfo {author} {\bibfnamefont {P.}~\bibnamefont {Li}}, \bibinfo {author} {\bibfnamefont {S.}~\bibnamefont {Portillo}}, \bibinfo {author} {\bibfnamefont {A.}~\bibnamefont {Fraser}}, \bibinfo {author} {\bibfnamefont {J.}~\bibnamefont {Burgess}}, \bibinfo {author} {\bibfnamefont {D.}~\bibnamefont {Fortin}}, \bibinfo {author} {\bibfnamefont {W.}~\bibnamefont {Hiebert}}, \ and\ \bibinfo {author} {\bibfnamefont {M.}~\bibnamefont {Freeman}},\ }\bibfield  {title} {\enquote {\bibinfo {title} {Nanomechanical torsional resonator torque magnetometry},}\ }\href {https://doi.org/10.1063/1.3540643} {\bibfield  {journal} {\bibinfo  {journal} {Journal of Applied Physics}\ }\textbf {\bibinfo {volume} {109}} (\bibinfo {year} {2011})}\BibitemShut {NoStop}%
\bibitem [{\citenamefont {Cong}\ \emph {et~al.}(2021)\citenamefont {Cong}, \citenamefont {Yuan}, \citenamefont {Bai}, \citenamefont {Wang}, \citenamefont {Zhao}, \citenamefont {Gao}, \citenamefont {Hu}, \citenamefont {Liu}, \citenamefont {Guo}, \citenamefont {Li} \emph {et~al.}}]{cong2021chip}%
  \BibitemOpen
  \bibfield  {author} {\bibinfo {author} {\bibfnamefont {L.}~\bibnamefont {Cong}}, \bibinfo {author} {\bibfnamefont {Z.}~\bibnamefont {Yuan}}, \bibinfo {author} {\bibfnamefont {Z.}~\bibnamefont {Bai}}, \bibinfo {author} {\bibfnamefont {X.}~\bibnamefont {Wang}}, \bibinfo {author} {\bibfnamefont {W.}~\bibnamefont {Zhao}}, \bibinfo {author} {\bibfnamefont {X.}~\bibnamefont {Gao}}, \bibinfo {author} {\bibfnamefont {X.}~\bibnamefont {Hu}}, \bibinfo {author} {\bibfnamefont {P.}~\bibnamefont {Liu}}, \bibinfo {author} {\bibfnamefont {W.}~\bibnamefont {Guo}}, \bibinfo {author} {\bibfnamefont {Q.}~\bibnamefont {Li}},  \emph {et~al.},\ }\bibfield  {title} {\enquote {\bibinfo {title} {On-chip torsion balances with femtonewton force resolution at room temperature enabled by carbon nanotube and graphene},}\ }\href {https://www.science.org/doi/abs/10.1126/sciadv.abd2358} {\bibfield  {journal} {\bibinfo  {journal} {Science Advances}\ }\textbf {\bibinfo {volume} {7}},\ \bibinfo {pages} {eabd2358} (\bibinfo {year}
  {2021})}\BibitemShut {NoStop}%
\bibitem [{\citenamefont {Kleiman}\ \emph {et~al.}(1985)\citenamefont {Kleiman}, \citenamefont {Kaminsky}, \citenamefont {Reppy}, \citenamefont {Pindak},\ and\ \citenamefont {Bishop}}]{kleiman1985single}%
  \BibitemOpen
  \bibfield  {author} {\bibinfo {author} {\bibfnamefont {R.}~\bibnamefont {Kleiman}}, \bibinfo {author} {\bibfnamefont {G.}~\bibnamefont {Kaminsky}}, \bibinfo {author} {\bibfnamefont {J.}~\bibnamefont {Reppy}}, \bibinfo {author} {\bibfnamefont {R.}~\bibnamefont {Pindak}}, \ and\ \bibinfo {author} {\bibfnamefont {D.}~\bibnamefont {Bishop}},\ }\bibfield  {title} {\enquote {\bibinfo {title} {Single-crystal silicon high-q torsional oscillators},}\ }\href {https://doi.org/10.1063/1.1138425} {\bibfield  {journal} {\bibinfo  {journal} {Review of scientific instruments}\ }\textbf {\bibinfo {volume} {56}},\ \bibinfo {pages} {2088--2091} (\bibinfo {year} {1985})}\BibitemShut {NoStop}%
\bibitem [{\citenamefont {Manley}\ \emph {et~al.}(2024)\citenamefont {Manley}, \citenamefont {Condos}, \citenamefont {Schlamminger}, \citenamefont {Pratt}, \citenamefont {Wilson},\ and\ \citenamefont {Terrano}}]{manley2024microscale}%
  \BibitemOpen
  \bibfield  {author} {\bibinfo {author} {\bibfnamefont {J.}~\bibnamefont {Manley}}, \bibinfo {author} {\bibfnamefont {C.}~\bibnamefont {Condos}}, \bibinfo {author} {\bibfnamefont {S.}~\bibnamefont {Schlamminger}}, \bibinfo {author} {\bibfnamefont {J.}~\bibnamefont {Pratt}}, \bibinfo {author} {\bibfnamefont {D.}~\bibnamefont {Wilson}}, \ and\ \bibinfo {author} {\bibfnamefont {W.}~\bibnamefont {Terrano}},\ }\bibfield  {title} {\enquote {\bibinfo {title} {Microscale torsion resonators for short-range gravity experiments},}\ }\href {https://doi.org/10.1103/PhysRevD.110.122005} {\bibfield  {journal} {\bibinfo  {journal} {Physical Review D}\ }\textbf {\bibinfo {volume} {110}},\ \bibinfo {pages} {122005} (\bibinfo {year} {2024})}\BibitemShut {NoStop}%
\bibitem [{\citenamefont {Chen}\ \emph {et~al.}(2016)\citenamefont {Chen}, \citenamefont {Tham}, \citenamefont {Krause}, \citenamefont {L\'opez}, \citenamefont {Fischbach},\ and\ \citenamefont {Decca}}]{PhysRevLett.116.221102}%
  \BibitemOpen
  \bibfield  {author} {\bibinfo {author} {\bibfnamefont {Y.-J.}\ \bibnamefont {Chen}}, \bibinfo {author} {\bibfnamefont {W.~K.}\ \bibnamefont {Tham}}, \bibinfo {author} {\bibfnamefont {D.~E.}\ \bibnamefont {Krause}}, \bibinfo {author} {\bibfnamefont {D.}~\bibnamefont {L\'opez}}, \bibinfo {author} {\bibfnamefont {E.}~\bibnamefont {Fischbach}}, \ and\ \bibinfo {author} {\bibfnamefont {R.~S.}\ \bibnamefont {Decca}},\ }\bibfield  {title} {\enquote {\bibinfo {title} {Stronger limits on hypothetical yukawa interactions in the 30--8000 nm range},}\ }\href {\doibase 10.1103/PhysRevLett.116.221102} {\bibfield  {journal} {\bibinfo  {journal} {Phys. Rev. Lett.}\ }\textbf {\bibinfo {volume} {116}},\ \bibinfo {pages} {221102} (\bibinfo {year} {2016})}\BibitemShut {NoStop}%
\bibitem [{\citenamefont {Agafonova}\ \emph {et~al.}(2024)\citenamefont {Agafonova}, \citenamefont {Rossello}, \citenamefont {Mekonnen},\ and\ \citenamefont {Hosten}}]{agafonova2024laser}%
  \BibitemOpen
  \bibfield  {author} {\bibinfo {author} {\bibfnamefont {S.}~\bibnamefont {Agafonova}}, \bibinfo {author} {\bibfnamefont {P.}~\bibnamefont {Rossello}}, \bibinfo {author} {\bibfnamefont {M.}~\bibnamefont {Mekonnen}}, \ and\ \bibinfo {author} {\bibfnamefont {O.}~\bibnamefont {Hosten}},\ }\bibfield  {title} {\enquote {\bibinfo {title} {Laser cooling a 1-milligram torsional pendulum to 240 microkelvins},}\ }\href {https://arxiv.org/abs/2408.09445} {\bibfield  {journal} {\bibinfo  {journal} {arXiv preprint arXiv:2408.09445}\ } (\bibinfo {year} {2024})}\BibitemShut {NoStop}%
\bibitem [{\citenamefont {Bose}\ \emph {et~al.}(2025)\citenamefont {Bose}, \citenamefont {Fuentes}, \citenamefont {Geraci}, \citenamefont {Khan}, \citenamefont {Qvarfort}, \citenamefont {Rademacher}, \citenamefont {Rashid}, \citenamefont {Toro{\v{s}}}, \citenamefont {Ulbricht},\ and\ \citenamefont {Wanjura}}]{bose2025massive}%
  \BibitemOpen
  \bibfield  {author} {\bibinfo {author} {\bibfnamefont {S.}~\bibnamefont {Bose}}, \bibinfo {author} {\bibfnamefont {I.}~\bibnamefont {Fuentes}}, \bibinfo {author} {\bibfnamefont {A.~A.}\ \bibnamefont {Geraci}}, \bibinfo {author} {\bibfnamefont {S.~M.}\ \bibnamefont {Khan}}, \bibinfo {author} {\bibfnamefont {S.}~\bibnamefont {Qvarfort}}, \bibinfo {author} {\bibfnamefont {M.}~\bibnamefont {Rademacher}}, \bibinfo {author} {\bibfnamefont {M.}~\bibnamefont {Rashid}}, \bibinfo {author} {\bibfnamefont {M.}~\bibnamefont {Toro{\v{s}}}}, \bibinfo {author} {\bibfnamefont {H.}~\bibnamefont {Ulbricht}}, \ and\ \bibinfo {author} {\bibfnamefont {C.~C.}\ \bibnamefont {Wanjura}},\ }\bibfield  {title} {\enquote {\bibinfo {title} {Massive quantum systems as interfaces of quantum mechanics and gravity},}\ }\href {https://journals.aps.org/rmp/abstract/10.1103/RevModPhys.97.015003} {\bibfield  {journal} {\bibinfo  {journal} {Rev. Mod. Phys.}\ }\textbf {\bibinfo {volume} {97}},\ \bibinfo {pages} {015003} (\bibinfo {year}
  {2025})}\BibitemShut {NoStop}%
\bibitem [{\citenamefont {Tsaturyan}\ \emph {et~al.}(2017)\citenamefont {Tsaturyan}, \citenamefont {Barg}, \citenamefont {Polzik},\ and\ \citenamefont {Schliesser}}]{tsaturyan2017ultracoherent}%
  \BibitemOpen
  \bibfield  {author} {\bibinfo {author} {\bibfnamefont {Y.}~\bibnamefont {Tsaturyan}}, \bibinfo {author} {\bibfnamefont {A.}~\bibnamefont {Barg}}, \bibinfo {author} {\bibfnamefont {E.~S.}\ \bibnamefont {Polzik}}, \ and\ \bibinfo {author} {\bibfnamefont {A.}~\bibnamefont {Schliesser}},\ }\bibfield  {title} {\enquote {\bibinfo {title} {Ultracoherent nanomechanical resonators via soft clamping and dissipation dilution},}\ }\href {https://doi.org/10.1038/nnano.2017.101} {\bibfield  {journal} {\bibinfo  {journal} {Nature Nanotechnology}\ }\textbf {\bibinfo {volume} {12}},\ \bibinfo {pages} {776--783} (\bibinfo {year} {2017})}\BibitemShut {NoStop}%
\bibitem [{\citenamefont {Engelsen}\ \emph {et~al.}(2024)\citenamefont {Engelsen}, \citenamefont {Beccari},\ and\ \citenamefont {Kippenberg}}]{engelsen2024ultrahigh}%
  \BibitemOpen
  \bibfield  {author} {\bibinfo {author} {\bibfnamefont {N.~J.}\ \bibnamefont {Engelsen}}, \bibinfo {author} {\bibfnamefont {A.}~\bibnamefont {Beccari}}, \ and\ \bibinfo {author} {\bibfnamefont {T.~J.}\ \bibnamefont {Kippenberg}},\ }\bibfield  {title} {\enquote {\bibinfo {title} {Ultrahigh-quality-factor micro-and nanomechanical resonators using dissipation dilution},}\ }\href {https://doi.org/10.1038/s41565-023-01597-8} {\bibfield  {journal} {\bibinfo  {journal} {Nature Nanotechnology}\ ,\ \bibinfo {pages} {1--13}} (\bibinfo {year} {2024})}\BibitemShut {NoStop}%
\bibitem [{\citenamefont {Ghadimi}\ \emph {et~al.}(2018)\citenamefont {Ghadimi}, \citenamefont {Fedorov}, \citenamefont {Engelsen}, \citenamefont {Bereyhi}, \citenamefont {Schilling}, \citenamefont {Wilson},\ and\ \citenamefont {Kippenberg}}]{ghadimi2018elastic}%
  \BibitemOpen
  \bibfield  {author} {\bibinfo {author} {\bibfnamefont {A.~H.}\ \bibnamefont {Ghadimi}}, \bibinfo {author} {\bibfnamefont {S.~A.}\ \bibnamefont {Fedorov}}, \bibinfo {author} {\bibfnamefont {N.~J.}\ \bibnamefont {Engelsen}}, \bibinfo {author} {\bibfnamefont {M.~J.}\ \bibnamefont {Bereyhi}}, \bibinfo {author} {\bibfnamefont {R.}~\bibnamefont {Schilling}}, \bibinfo {author} {\bibfnamefont {D.~J.}\ \bibnamefont {Wilson}}, \ and\ \bibinfo {author} {\bibfnamefont {T.~J.}\ \bibnamefont {Kippenberg}},\ }\bibfield  {title} {\enquote {\bibinfo {title} {Elastic strain engineering for ultralow mechanical dissipation},}\ }\href {https://science.sciencemag.org/content/360/6390/764.extract} {\bibfield  {journal} {\bibinfo  {journal} {Science}\ }\textbf {\bibinfo {volume} {360}},\ \bibinfo {pages} {764--768} (\bibinfo {year} {2018})}\BibitemShut {NoStop}%
\bibitem [{\citenamefont {Bereyhi}\ \emph {et~al.}(2022)\citenamefont {Bereyhi}, \citenamefont {Beccari}, \citenamefont {Groth}, \citenamefont {Fedorov}, \citenamefont {Arabmoheghi}, \citenamefont {Kippenberg},\ and\ \citenamefont {Engelsen}}]{bereyhi2022hierarchical}%
  \BibitemOpen
  \bibfield  {author} {\bibinfo {author} {\bibfnamefont {M.~J.}\ \bibnamefont {Bereyhi}}, \bibinfo {author} {\bibfnamefont {A.}~\bibnamefont {Beccari}}, \bibinfo {author} {\bibfnamefont {R.}~\bibnamefont {Groth}}, \bibinfo {author} {\bibfnamefont {S.~A.}\ \bibnamefont {Fedorov}}, \bibinfo {author} {\bibfnamefont {A.}~\bibnamefont {Arabmoheghi}}, \bibinfo {author} {\bibfnamefont {T.~J.}\ \bibnamefont {Kippenberg}}, \ and\ \bibinfo {author} {\bibfnamefont {N.~J.}\ \bibnamefont {Engelsen}},\ }\bibfield  {title} {\enquote {\bibinfo {title} {Hierarchical tensile structures with ultralow mechanical dissipation},}\ }\href {https://doi.org/10.1038/s41467-022-30586-z} {\bibfield  {journal} {\bibinfo  {journal} {Nature Communications}\ }\textbf {\bibinfo {volume} {13}},\ \bibinfo {pages} {3097} (\bibinfo {year} {2022})}\BibitemShut {NoStop}%
\bibitem [{\citenamefont {Buckley}(1914)}]{buckley1914lxxxiv}%
  \BibitemOpen
  \bibfield  {author} {\bibinfo {author} {\bibfnamefont {J.}~\bibnamefont {Buckley}},\ }\bibfield  {title} {\enquote {\bibinfo {title} {Lxxxiv. the bifilar property of twisted strips},}\ }\href {https://doi.org/10.1080/14786441208635264} {\bibfield  {journal} {\bibinfo  {journal} {The London, Edinburgh, and Dublin Philosophical Magazine and Journal of Science}\ }\textbf {\bibinfo {volume} {28}},\ \bibinfo {pages} {778--787} (\bibinfo {year} {1914})}\BibitemShut {NoStop}%
\bibitem [{\citenamefont {Barr{\'e} Saint-Venant}(1855)}]{barre1855memoire}%
  \BibitemOpen
  \bibfield  {author} {\bibinfo {author} {\bibfnamefont {A.}~\bibnamefont {Barr{\'e} Saint-Venant}},\ }\bibfield  {title} {\enquote {\bibinfo {title} {M{\'e}moire sur la torsion des prismes,[essay on twisting prisms], m{\'e}moires des savants {\'e}trangers [essays of foreign scholars]},}\ }\href@noop {} {\bibfield  {journal} {\bibinfo  {journal} {CR Acad. Sci}\ }\textbf {\bibinfo {volume} {14}},\ \bibinfo {pages} {233--560} (\bibinfo {year} {1855})}\BibitemShut {NoStop}%
\bibitem [{\citenamefont {Fedorov}\ \emph {et~al.}(2019)\citenamefont {Fedorov}, \citenamefont {Engelsen}, \citenamefont {Ghadimi}, \citenamefont {Bereyhi}, \citenamefont {Schilling}, \citenamefont {Wilson},\ and\ \citenamefont {Kippenberg}}]{fedorov2019generalized}%
  \BibitemOpen
  \bibfield  {author} {\bibinfo {author} {\bibfnamefont {S.~A.}\ \bibnamefont {Fedorov}}, \bibinfo {author} {\bibfnamefont {N.~J.}\ \bibnamefont {Engelsen}}, \bibinfo {author} {\bibfnamefont {A.~H.}\ \bibnamefont {Ghadimi}}, \bibinfo {author} {\bibfnamefont {M.~J.}\ \bibnamefont {Bereyhi}}, \bibinfo {author} {\bibfnamefont {R.}~\bibnamefont {Schilling}}, \bibinfo {author} {\bibfnamefont {D.~J.}\ \bibnamefont {Wilson}}, \ and\ \bibinfo {author} {\bibfnamefont {T.~J.}\ \bibnamefont {Kippenberg}},\ }\bibfield  {title} {\enquote {\bibinfo {title} {Generalized dissipation dilution in strained mechanical resonators},}\ }\href {https://doi.org/10.1103/PhysRevB.99.054107} {\bibfield  {journal} {\bibinfo  {journal} {Physical Review B}\ }\textbf {\bibinfo {volume} {99}},\ \bibinfo {pages} {054107} (\bibinfo {year} {2019})}\BibitemShut {NoStop}%
\bibitem [{\citenamefont {Villanueva}\ and\ \citenamefont {Schmid}(2014)}]{villanueva2014evidence}%
  \BibitemOpen
  \bibfield  {author} {\bibinfo {author} {\bibfnamefont {L.~G.}\ \bibnamefont {Villanueva}}\ and\ \bibinfo {author} {\bibfnamefont {S.}~\bibnamefont {Schmid}},\ }\bibfield  {title} {\enquote {\bibinfo {title} {Evidence of surface loss as ubiquitous limiting damping mechanism in sin micro-and nanomechanical resonators},}\ }\href {https://doi.org/10.1103/PhysRevLett.113.227201} {\bibfield  {journal} {\bibinfo  {journal} {Physical Review Letters}\ }\textbf {\bibinfo {volume} {113}},\ \bibinfo {pages} {227201} (\bibinfo {year} {2014})}\BibitemShut {NoStop}%
\bibitem [{\citenamefont {Bereyhi}\ \emph {et~al.}(2019)\citenamefont {Bereyhi}, \citenamefont {Beccari}, \citenamefont {Fedorov}, \citenamefont {Ghadimi}, \citenamefont {Schilling}, \citenamefont {Wilson}, \citenamefont {Engelsen},\ and\ \citenamefont {Kippenberg}}]{bereyhi2019clamp}%
  \BibitemOpen
  \bibfield  {author} {\bibinfo {author} {\bibfnamefont {M.~J.}\ \bibnamefont {Bereyhi}}, \bibinfo {author} {\bibfnamefont {A.}~\bibnamefont {Beccari}}, \bibinfo {author} {\bibfnamefont {S.~A.}\ \bibnamefont {Fedorov}}, \bibinfo {author} {\bibfnamefont {A.~H.}\ \bibnamefont {Ghadimi}}, \bibinfo {author} {\bibfnamefont {R.}~\bibnamefont {Schilling}}, \bibinfo {author} {\bibfnamefont {D.~J.}\ \bibnamefont {Wilson}}, \bibinfo {author} {\bibfnamefont {N.~J.}\ \bibnamefont {Engelsen}}, \ and\ \bibinfo {author} {\bibfnamefont {T.~J.}\ \bibnamefont {Kippenberg}},\ }\bibfield  {title} {\enquote {\bibinfo {title} {Clamp-tapering increases the quality factor of stressed nanobeams},}\ }\href {https://pubs.acs.org/doi/abs/10.1021/acs.nanolett.8b04942} {\bibfield  {journal} {\bibinfo  {journal} {Nano letters}\ }\textbf {\bibinfo {volume} {19}},\ \bibinfo {pages} {2329--2333} (\bibinfo {year} {2019})}\BibitemShut {NoStop}%
\bibitem [{\citenamefont {Sadeghi}\ \emph {et~al.}(2019)\citenamefont {Sadeghi}, \citenamefont {Tanzer}, \citenamefont {Christensen},\ and\ \citenamefont {Schmid}}]{sadeghi2019influence}%
  \BibitemOpen
  \bibfield  {author} {\bibinfo {author} {\bibfnamefont {P.}~\bibnamefont {Sadeghi}}, \bibinfo {author} {\bibfnamefont {M.}~\bibnamefont {Tanzer}}, \bibinfo {author} {\bibfnamefont {S.~L.}\ \bibnamefont {Christensen}}, \ and\ \bibinfo {author} {\bibfnamefont {S.}~\bibnamefont {Schmid}},\ }\bibfield  {title} {\enquote {\bibinfo {title} {Influence of clamp-widening on the quality factor of nanomechanical silicon nitride resonators},}\ }\href {https://doi.org/10.1063/1.5111712} {\bibfield  {journal} {\bibinfo  {journal} {Journal of Applied Physics}\ }\textbf {\bibinfo {volume} {126}} (\bibinfo {year} {2019})}\BibitemShut {NoStop}%
\bibitem [{\citenamefont {Norte}\ \emph {et~al.}(2016)\citenamefont {Norte}, \citenamefont {Moura},\ and\ \citenamefont {Gr{\"o}blacher}}]{norte2016mechanical}%
  \BibitemOpen
  \bibfield  {author} {\bibinfo {author} {\bibfnamefont {R.~A.}\ \bibnamefont {Norte}}, \bibinfo {author} {\bibfnamefont {J.~P.}\ \bibnamefont {Moura}}, \ and\ \bibinfo {author} {\bibfnamefont {S.}~\bibnamefont {Gr{\"o}blacher}},\ }\bibfield  {title} {\enquote {\bibinfo {title} {Mechanical resonators for quantum optomechanics experiments at room temperature},}\ }\href {https://journals.aps.org/prl/abstract/10.1103/PhysRevLett.116.147202} {\bibfield  {journal} {\bibinfo  {journal} {Physical Review Letters}\ }\textbf {\bibinfo {volume} {116}},\ \bibinfo {pages} {147202} (\bibinfo {year} {2016})}\BibitemShut {NoStop}%
\bibitem [{\citenamefont {Reinhardt}\ \emph {et~al.}(2016)\citenamefont {Reinhardt}, \citenamefont {M{\"u}ller}, \citenamefont {Bourassa},\ and\ \citenamefont {Sankey}}]{reinhardt2016ultralow}%
  \BibitemOpen
  \bibfield  {author} {\bibinfo {author} {\bibfnamefont {C.}~\bibnamefont {Reinhardt}}, \bibinfo {author} {\bibfnamefont {T.}~\bibnamefont {M{\"u}ller}}, \bibinfo {author} {\bibfnamefont {A.}~\bibnamefont {Bourassa}}, \ and\ \bibinfo {author} {\bibfnamefont {J.~C.}\ \bibnamefont {Sankey}},\ }\bibfield  {title} {\enquote {\bibinfo {title} {Ultralow-noise sin trampoline resonators for sensing and optomechanics},}\ }\href {https://journals.aps.org/prx/abstract/10.1103/PhysRevX.6.021001} {\bibfield  {journal} {\bibinfo  {journal} {Physical Review X}\ }\textbf {\bibinfo {volume} {6}},\ \bibinfo {pages} {021001} (\bibinfo {year} {2016})}\BibitemShut {NoStop}%
\bibitem [{\citenamefont {Yu}\ \emph {et~al.}(2012)\citenamefont {Yu}, \citenamefont {Purdy},\ and\ \citenamefont {Regal}}]{yu2012control}%
  \BibitemOpen
  \bibfield  {author} {\bibinfo {author} {\bibfnamefont {P.-L.}\ \bibnamefont {Yu}}, \bibinfo {author} {\bibfnamefont {T.}~\bibnamefont {Purdy}}, \ and\ \bibinfo {author} {\bibfnamefont {C.}~\bibnamefont {Regal}},\ }\bibfield  {title} {\enquote {\bibinfo {title} {Control of material damping in high-q membrane microresonators},}\ }\href@noop {} {\bibfield  {journal} {\bibinfo  {journal} {Physical Review Letters}\ }\textbf {\bibinfo {volume} {108}},\ \bibinfo {pages} {083603} (\bibinfo {year} {2012})}\BibitemShut {NoStop}%
\bibitem [{\citenamefont {Ghadimi}(2018)}]{ghadimi2018ultra}%
  \BibitemOpen
  \bibfield  {author} {\bibinfo {author} {\bibfnamefont {A.~H.}\ \bibnamefont {Ghadimi}},\ }\bibfield  {title} {\enquote {\bibinfo {title} {Ultra-coherent nano-mechanical resonators for quantum optomechanics at room temperature},}\ }\href {https://infoscience.epfl.ch/entities/publication/9a905d97-69f1-4dfd-a2c2-ded12e04ef23} {\  (\bibinfo {year} {2018})}\BibitemShut {NoStop}%
\bibitem [{COM()}]{COMSOLuserguide}%
  \BibitemOpen
  \href {https://doc.comsol.com/5.4/doc/com.comsol.help.sme/StructuralMechanicsModuleUsersGuide.pdf} {\emph {\bibinfo {title} {COMSOL v5.4 Structural Mechanics Module User's Guide}}}\BibitemShut {NoStop}%
\bibitem [{\citenamefont {Bendsoe}\ and\ \citenamefont {Sigmund}(2013)}]{bendsoe2013topology}%
  \BibitemOpen
  \bibfield  {author} {\bibinfo {author} {\bibfnamefont {M.~P.}\ \bibnamefont {Bendsoe}}\ and\ \bibinfo {author} {\bibfnamefont {O.}~\bibnamefont {Sigmund}},\ }\bibfield  {title} {\enquote {\bibinfo {title} {Topology optimization: Theory, methods, and applications},}\ }\href {https://link.springer.com/book/10.1007/978-3-662-05086-6} {\  (\bibinfo {year} {2013})}\BibitemShut {NoStop}%
\bibitem [{\citenamefont {Mockus}\ and\ \citenamefont {Mockus}(1991)}]{mockus1991bayesian}%
  \BibitemOpen
  \bibfield  {author} {\bibinfo {author} {\bibfnamefont {J.~B.}\ \bibnamefont {Mockus}}\ and\ \bibinfo {author} {\bibfnamefont {L.~J.}\ \bibnamefont {Mockus}},\ }\bibfield  {title} {\enquote {\bibinfo {title} {Bayesian approach to global optimization and application to multiobjective and constrained problems},}\ }\href {https://link.springer.com/article/10.1007/BF00940509} {\bibfield  {journal} {\bibinfo  {journal} {Journal of optimization theory and applications}\ }\textbf {\bibinfo {volume} {70}},\ \bibinfo {pages} {157--172} (\bibinfo {year} {1991})}\BibitemShut {NoStop}%
\bibitem [{\citenamefont {Snoek}\ \emph {et~al.}(2012)\citenamefont {Snoek}, \citenamefont {Larochelle},\ and\ \citenamefont {Adams}}]{snoek2012practical}%
  \BibitemOpen
  \bibfield  {author} {\bibinfo {author} {\bibfnamefont {J.}~\bibnamefont {Snoek}}, \bibinfo {author} {\bibfnamefont {H.}~\bibnamefont {Larochelle}}, \ and\ \bibinfo {author} {\bibfnamefont {R.~P.}\ \bibnamefont {Adams}},\ }\bibfield  {title} {\enquote {\bibinfo {title} {Practical bayesian optimization of machine learning algorithms},}\ }\href {https://proceedings.neurips.cc/paper/2012/hash/05311655a15b75fab86956663e1819cd-Abstract.html} {\bibfield  {journal} {\bibinfo  {journal} {Advances in neural information processing systems}\ }\textbf {\bibinfo {volume} {25}} (\bibinfo {year} {2012})}\BibitemShut {NoStop}%
\bibitem [{\citenamefont {H{\o}j}\ \emph {et~al.}(2021)\citenamefont {H{\o}j}, \citenamefont {Wang}, \citenamefont {Gao}, \citenamefont {Hoff}, \citenamefont {Sigmund},\ and\ \citenamefont {Andersen}}]{hoj2021ultra}%
  \BibitemOpen
  \bibfield  {author} {\bibinfo {author} {\bibfnamefont {D.}~\bibnamefont {H{\o}j}}, \bibinfo {author} {\bibfnamefont {F.}~\bibnamefont {Wang}}, \bibinfo {author} {\bibfnamefont {W.}~\bibnamefont {Gao}}, \bibinfo {author} {\bibfnamefont {U.~B.}\ \bibnamefont {Hoff}}, \bibinfo {author} {\bibfnamefont {O.}~\bibnamefont {Sigmund}}, \ and\ \bibinfo {author} {\bibfnamefont {U.~L.}\ \bibnamefont {Andersen}},\ }\bibfield  {title} {\enquote {\bibinfo {title} {Ultra-coherent nanomechanical resonators based on inverse design},}\ }\href {https://doi.org/10.1038/s41565-023-01597-8} {\bibfield  {journal} {\bibinfo  {journal} {Nature communications}\ }\textbf {\bibinfo {volume} {12}},\ \bibinfo {pages} {5766} (\bibinfo {year} {2021})}\BibitemShut {NoStop}%
\bibitem [{\citenamefont {Norder}\ \emph {et~al.}(2024)\citenamefont {Norder}, \citenamefont {Yin}, \citenamefont {de~Jong}, \citenamefont {Stallone}, \citenamefont {Aydogmus}, \citenamefont {Sberna}, \citenamefont {Bessa},\ and\ \citenamefont {Norte}}]{norder2024pentagonal}%
  \BibitemOpen
  \bibfield  {author} {\bibinfo {author} {\bibfnamefont {L.}~\bibnamefont {Norder}}, \bibinfo {author} {\bibfnamefont {S.}~\bibnamefont {Yin}}, \bibinfo {author} {\bibfnamefont {M.}~\bibnamefont {de~Jong}}, \bibinfo {author} {\bibfnamefont {F.}~\bibnamefont {Stallone}}, \bibinfo {author} {\bibfnamefont {H.}~\bibnamefont {Aydogmus}}, \bibinfo {author} {\bibfnamefont {P.}~\bibnamefont {Sberna}}, \bibinfo {author} {\bibfnamefont {M.}~\bibnamefont {Bessa}}, \ and\ \bibinfo {author} {\bibfnamefont {R.}~\bibnamefont {Norte}},\ }\bibfield  {title} {\enquote {\bibinfo {title} {Pentagonal photonic crystal mirrors: Scalable lightsails with enhanced acceleration via neural topology optimization},}\ }\href {https://doi.org/10.1038/s41467-025-57749-y} {\bibfield  {journal} {\bibinfo  {journal} {arXiv preprint arXiv:2407.07896}\ } (\bibinfo {year} {2024})}\BibitemShut {NoStop}%
\bibitem [{\citenamefont {Shin}\ \emph {et~al.}(2022)\citenamefont {Shin}, \citenamefont {Cupertino}, \citenamefont {de~Jong}, \citenamefont {Steeneken}, \citenamefont {Bessa},\ and\ \citenamefont {Norte}}]{shin2022spiderweb}%
  \BibitemOpen
  \bibfield  {author} {\bibinfo {author} {\bibfnamefont {D.}~\bibnamefont {Shin}}, \bibinfo {author} {\bibfnamefont {A.}~\bibnamefont {Cupertino}}, \bibinfo {author} {\bibfnamefont {M.~H.}\ \bibnamefont {de~Jong}}, \bibinfo {author} {\bibfnamefont {P.~G.}\ \bibnamefont {Steeneken}}, \bibinfo {author} {\bibfnamefont {M.~A.}\ \bibnamefont {Bessa}}, \ and\ \bibinfo {author} {\bibfnamefont {R.~A.}\ \bibnamefont {Norte}},\ }\bibfield  {title} {\enquote {\bibinfo {title} {Spiderweb nanomechanical resonators via bayesian optimization: inspired by nature and guided by machine learning},}\ }\href {https://doi.org/10.1002/adma.202106248} {\bibfield  {journal} {\bibinfo  {journal} {Advanced Materials}\ }\textbf {\bibinfo {volume} {34}},\ \bibinfo {pages} {2106248} (\bibinfo {year} {2022})}\BibitemShut {NoStop}%
\bibitem [{\citenamefont {Cupertino}\ \emph {et~al.}(2024)\citenamefont {Cupertino}, \citenamefont {Shin}, \citenamefont {Guo}, \citenamefont {Steeneken}, \citenamefont {Bessa},\ and\ \citenamefont {Norte}}]{cupertino2024centimeter}%
  \BibitemOpen
  \bibfield  {author} {\bibinfo {author} {\bibfnamefont {A.}~\bibnamefont {Cupertino}}, \bibinfo {author} {\bibfnamefont {D.}~\bibnamefont {Shin}}, \bibinfo {author} {\bibfnamefont {L.}~\bibnamefont {Guo}}, \bibinfo {author} {\bibfnamefont {P.~G.}\ \bibnamefont {Steeneken}}, \bibinfo {author} {\bibfnamefont {M.~A.}\ \bibnamefont {Bessa}}, \ and\ \bibinfo {author} {\bibfnamefont {R.~A.}\ \bibnamefont {Norte}},\ }\bibfield  {title} {\enquote {\bibinfo {title} {Centimeter-scale nanomechanical resonators with low dissipation},}\ }\href {https://doi.org/10.1038/s41467-024-48183-7} {\bibfield  {journal} {\bibinfo  {journal} {Nature Communications}\ }\textbf {\bibinfo {volume} {15}},\ \bibinfo {pages} {4255} (\bibinfo {year} {2024})}\BibitemShut {NoStop}%
\bibitem [{\citenamefont {Pardalos}\ \emph {et~al.}(2021)\citenamefont {Pardalos}, \citenamefont {Rasskazova}, \citenamefont {Vrahatis} \emph {et~al.}}]{pardalos2021black}%
  \BibitemOpen
  \bibfield  {author} {\bibinfo {author} {\bibfnamefont {P.~M.}\ \bibnamefont {Pardalos}}, \bibinfo {author} {\bibfnamefont {V.}~\bibnamefont {Rasskazova}}, \bibinfo {author} {\bibfnamefont {M.~N.}\ \bibnamefont {Vrahatis}},  \emph {et~al.},\ }\bibfield  {title} {\enquote {\bibinfo {title} {Black box optimization, machine learning, and no-free lunch theorems},}\ }\href {https://link.springer.com/book/10.1007/978-3-030-66515-9} {\ \textbf {\bibinfo {volume} {170}} (\bibinfo {year} {2021})}\BibitemShut {NoStop}%
\bibitem [{\citenamefont {Wang}(2023)}]{wang2023intuitive}%
  \BibitemOpen
  \bibfield  {author} {\bibinfo {author} {\bibfnamefont {J.}~\bibnamefont {Wang}},\ }\bibfield  {title} {\enquote {\bibinfo {title} {An intuitive tutorial to gaussian processes regression},}\ }\href {https://ieeexplore.ieee.org/abstract/document/10360364} {\bibfield  {journal} {\bibinfo  {journal} {Computing in Science \& Engineering}\ } (\bibinfo {year} {2023})}\BibitemShut {NoStop}%
\bibitem [{\citenamefont {Gan}\ \emph {et~al.}(2021)\citenamefont {Gan}, \citenamefont {Ji},\ and\ \citenamefont {Liang}}]{gan2021acquisition}%
  \BibitemOpen
  \bibfield  {author} {\bibinfo {author} {\bibfnamefont {W.}~\bibnamefont {Gan}}, \bibinfo {author} {\bibfnamefont {Z.}~\bibnamefont {Ji}}, \ and\ \bibinfo {author} {\bibfnamefont {Y.}~\bibnamefont {Liang}},\ }\bibfield  {title} {\enquote {\bibinfo {title} {Acquisition functions in bayesian optimization},}\ }\href {https://ieeexplore.ieee.org/document/9696089} {\ ,\ \bibinfo {pages} {129--135} (\bibinfo {year} {2021})}\BibitemShut {NoStop}%
\bibitem [{\citenamefont {Hoffman}\ \emph {et~al.}(2011)\citenamefont {Hoffman}, \citenamefont {Brochu}, \citenamefont {De~Freitas} \emph {et~al.}}]{hoffman2011portfolio}%
  \BibitemOpen
  \bibfield  {author} {\bibinfo {author} {\bibfnamefont {M.}~\bibnamefont {Hoffman}}, \bibinfo {author} {\bibfnamefont {E.}~\bibnamefont {Brochu}}, \bibinfo {author} {\bibfnamefont {N.}~\bibnamefont {De~Freitas}},  \emph {et~al.},\ }\bibfield  {title} {\enquote {\bibinfo {title} {Portfolio allocation for bayesian optimization.}}\ }\bibfield  {booktitle} {\emph {\bibinfo {booktitle} {UAI}},\ }\href {https://dl.acm.org/doi/10.5555/3020548.3020587} {\ ,\ \bibinfo {pages} {327--336} (\bibinfo {year} {2011})}\BibitemShut {NoStop}%
\bibitem [{\citenamefont {Jones}\ \emph {et~al.}(1998)\citenamefont {Jones}, \citenamefont {Schonlau},\ and\ \citenamefont {Welch}}]{jones1998expensive}%
  \BibitemOpen
  \bibfield  {author} {\bibinfo {author} {\bibfnamefont {D.~R.}\ \bibnamefont {Jones}}, \bibinfo {author} {\bibfnamefont {M.}~\bibnamefont {Schonlau}}, \ and\ \bibinfo {author} {\bibfnamefont {W.~J.}\ \bibnamefont {Welch}},\ }\bibfield  {title} {\enquote {\bibinfo {title} {Efficient global optimization of expensive blackbox functions},}\ }\href {https://link.springer.com/article/10.1023/A:1008306431147} {\bibfield  {journal} {\bibinfo  {journal} {Journal of Global Optimization}\ }\textbf {\bibinfo {volume} {13}},\ \bibinfo {pages} {455--492} (\bibinfo {year} {1998})}\BibitemShut {NoStop}%
\bibitem [{Mul()}]{MultipleParameter}%
  \BibitemOpen
  \href@noop {} {}\bibinfo {note} {Note that while our analysis is limited to one or two parameters, the Bayesian optimization algorithm can easily handle more, at the expense of increased runtime~\cite{pardalos2021black}}\BibitemShut {NoStop}%
\bibitem [{\citenamefont {Agrawal}(2025)}]{agrawal2025ultra}%
  \BibitemOpen
  \bibfield  {author} {\bibinfo {author} {\bibfnamefont {A.~R.}\ \bibnamefont {Agrawal}},\ }\bibfield  {title} {\enquote {\bibinfo {title} {Ultra-high-q membrane optomechanics with a twist},}\ }\href {https://repository.arizona.edu/handle/10150/675776} {\  (\bibinfo {year} {2025})}\BibitemShut {NoStop}%
\bibitem [{\citenamefont {Pluchar}\ \emph {et~al.}(2025)\citenamefont {Pluchar}, \citenamefont {Agrawal},\ and\ \citenamefont {Wilson}}]{pluchar2025quantum}%
  \BibitemOpen
  \bibfield  {author} {\bibinfo {author} {\bibfnamefont {C.~M.}\ \bibnamefont {Pluchar}}, \bibinfo {author} {\bibfnamefont {A.~R.}\ \bibnamefont {Agrawal}}, \ and\ \bibinfo {author} {\bibfnamefont {D.~J.}\ \bibnamefont {Wilson}},\ }\bibfield  {title} {\enquote {\bibinfo {title} {Quantum-limited optical lever measurement of a torsion oscillator},}\ }\href {https://opg.optica.org/optica/abstract.cfm?URI=optica-12-3-418} {\bibfield  {journal} {\bibinfo  {journal} {Optica}\ }\textbf {\bibinfo {volume} {12}},\ \bibinfo {pages} {418--423} (\bibinfo {year} {2025})}\BibitemShut {NoStop}%
\bibitem [{\citenamefont {Blom}\ \emph {et~al.}(1992)\citenamefont {Blom}, \citenamefont {Bouwstra}, \citenamefont {Elwenspoek},\ and\ \citenamefont {Fluitman}}]{blom1992dependence}%
  \BibitemOpen
  \bibfield  {author} {\bibinfo {author} {\bibfnamefont {F.}~\bibnamefont {Blom}}, \bibinfo {author} {\bibfnamefont {S.}~\bibnamefont {Bouwstra}}, \bibinfo {author} {\bibfnamefont {M.}~\bibnamefont {Elwenspoek}}, \ and\ \bibinfo {author} {\bibfnamefont {J.}~\bibnamefont {Fluitman}},\ }\bibfield  {title} {\enquote {\bibinfo {title} {Dependence of the quality factor of micromachined silicon beam resonators on pressure and geometry},}\ }\href {https://doi-org.ezproxy4.library.arizona.edu/10.1116/1.586300} {\bibfield  {journal} {\bibinfo  {journal} {Journal of Vacuum Science \& Technology B: Microelectronics and Nanometer Structures Processing, Measurement, and Phenomena}\ }\textbf {\bibinfo {volume} {10}},\ \bibinfo {pages} {19--26} (\bibinfo {year} {1992})}\BibitemShut {NoStop}%
\bibitem [{\citenamefont {Wilson}\ \emph {et~al.}(2009)\citenamefont {Wilson}, \citenamefont {Regal}, \citenamefont {Papp},\ and\ \citenamefont {Kimble}}]{wilson2009cavity}%
  \BibitemOpen
  \bibfield  {author} {\bibinfo {author} {\bibfnamefont {D.~J.}\ \bibnamefont {Wilson}}, \bibinfo {author} {\bibfnamefont {C.~A.}\ \bibnamefont {Regal}}, \bibinfo {author} {\bibfnamefont {S.~B.}\ \bibnamefont {Papp}}, \ and\ \bibinfo {author} {\bibfnamefont {H.}~\bibnamefont {Kimble}},\ }\bibfield  {title} {\enquote {\bibinfo {title} {Cavity optomechanics with stoichiometric sin films},}\ }\href {https://journals.aps.org/prl/abstract/10.1103/PhysRevLett.103.207204} {\bibfield  {journal} {\bibinfo  {journal} {Physical Review Letters}\ }\textbf {\bibinfo {volume} {103}},\ \bibinfo {pages} {207204} (\bibinfo {year} {2009})}\BibitemShut {NoStop}%
\bibitem [{\citenamefont {Gisler}\ \emph {et~al.}(2022)\citenamefont {Gisler}, \citenamefont {Helal}, \citenamefont {Sabonis}, \citenamefont {Grob}, \citenamefont {H{\'e}ritier}, \citenamefont {Degen}, \citenamefont {Ghadimi},\ and\ \citenamefont {Eichler}}]{gisler2022soft}%
  \BibitemOpen
  \bibfield  {author} {\bibinfo {author} {\bibfnamefont {T.}~\bibnamefont {Gisler}}, \bibinfo {author} {\bibfnamefont {M.}~\bibnamefont {Helal}}, \bibinfo {author} {\bibfnamefont {D.}~\bibnamefont {Sabonis}}, \bibinfo {author} {\bibfnamefont {U.}~\bibnamefont {Grob}}, \bibinfo {author} {\bibfnamefont {M.}~\bibnamefont {H{\'e}ritier}}, \bibinfo {author} {\bibfnamefont {C.~L.}\ \bibnamefont {Degen}}, \bibinfo {author} {\bibfnamefont {A.~H.}\ \bibnamefont {Ghadimi}}, \ and\ \bibinfo {author} {\bibfnamefont {A.}~\bibnamefont {Eichler}},\ }\bibfield  {title} {\enquote {\bibinfo {title} {Soft-clamped silicon nitride string resonators at millikelvin temperatures},}\ }\href {https://doi.org/10.1103/PhysRevLett.129.104301} {\bibfield  {journal} {\bibinfo  {journal} {Physical Review Letters}\ }\textbf {\bibinfo {volume} {129}},\ \bibinfo {pages} {104301} (\bibinfo {year} {2022})}\BibitemShut {NoStop}%
\bibitem [{\citenamefont {Yuan}\ \emph {et~al.}(2015)\citenamefont {Yuan}, \citenamefont {Cohen},\ and\ \citenamefont {Steele}}]{yuan2015silicon}%
  \BibitemOpen
  \bibfield  {author} {\bibinfo {author} {\bibfnamefont {M.}~\bibnamefont {Yuan}}, \bibinfo {author} {\bibfnamefont {M.~A.}\ \bibnamefont {Cohen}}, \ and\ \bibinfo {author} {\bibfnamefont {G.~A.}\ \bibnamefont {Steele}},\ }\bibfield  {title} {\enquote {\bibinfo {title} {Silicon nitride membrane resonators at millikelvin temperatures with quality factors exceeding 108},}\ }\href {https://doi.org/10.1063/1.4938747} {\bibfield  {journal} {\bibinfo  {journal} {Applied Physics Letters}\ }\textbf {\bibinfo {volume} {107}} (\bibinfo {year} {2015})}\BibitemShut {NoStop}%
\bibitem [{\citenamefont {Fischer}\ \emph {et~al.}(2016)\citenamefont {Fischer}, \citenamefont {Kampel}, \citenamefont {Assump{\c{c}}{\~a}o}, \citenamefont {Yu}, \citenamefont {Cicak}, \citenamefont {Peterson}, \citenamefont {Simmonds},\ and\ \citenamefont {Regal}}]{fischer2016optical}%
  \BibitemOpen
  \bibfield  {author} {\bibinfo {author} {\bibfnamefont {R.}~\bibnamefont {Fischer}}, \bibinfo {author} {\bibfnamefont {N.}~\bibnamefont {Kampel}}, \bibinfo {author} {\bibfnamefont {G.}~\bibnamefont {Assump{\c{c}}{\~a}o}}, \bibinfo {author} {\bibfnamefont {P.-L.}\ \bibnamefont {Yu}}, \bibinfo {author} {\bibfnamefont {K.}~\bibnamefont {Cicak}}, \bibinfo {author} {\bibfnamefont {R.}~\bibnamefont {Peterson}}, \bibinfo {author} {\bibfnamefont {R.}~\bibnamefont {Simmonds}}, \ and\ \bibinfo {author} {\bibfnamefont {C.}~\bibnamefont {Regal}},\ }\bibfield  {title} {\enquote {\bibinfo {title} {Optical probing of mechanical loss of a si$_3$n$_4$ membrane below 100 mk},}\ }\href {https://arxiv.org/abs/1611.00878} {\bibfield  {journal} {\bibinfo  {journal} {arXiv preprint arXiv:1611.00878}\ } (\bibinfo {year} {2016})}\BibitemShut {NoStop}%
\bibitem [{\citenamefont {Shin}\ \emph {et~al.}(2025)\citenamefont {Shin}, \citenamefont {Hayward}, \citenamefont {Fife}, \citenamefont {Menon},\ and\ \citenamefont {Sudhir}}]{shin2024laser}%
  \BibitemOpen
  \bibfield  {author} {\bibinfo {author} {\bibfnamefont {D.-C.}\ \bibnamefont {Shin}}, \bibinfo {author} {\bibfnamefont {T.~M.}\ \bibnamefont {Hayward}}, \bibinfo {author} {\bibfnamefont {D.}~\bibnamefont {Fife}}, \bibinfo {author} {\bibfnamefont {R.}~\bibnamefont {Menon}}, \ and\ \bibinfo {author} {\bibfnamefont {V.}~\bibnamefont {Sudhir}},\ }\bibfield  {title} {\enquote {\bibinfo {title} {Active laser cooling of a centimeter-scale torsional oscillator},}\ }\href {https://opg.optica.org/abstract.cfm?uri=optica-12-4-473} {\bibfield  {journal} {\bibinfo  {journal} {Optica}\ }\textbf {\bibinfo {volume} {12}},\ \bibinfo {pages} {473--478} (\bibinfo {year} {2025})}\BibitemShut {NoStop}%
\bibitem [{\citenamefont {Hao}\ and\ \citenamefont {Purdy}(2024)}]{hao2024back}%
  \BibitemOpen
  \bibfield  {author} {\bibinfo {author} {\bibfnamefont {S.}~\bibnamefont {Hao}}\ and\ \bibinfo {author} {\bibfnamefont {T.~P.}\ \bibnamefont {Purdy}},\ }\bibfield  {title} {\enquote {\bibinfo {title} {Back action evasion in optical lever detection},}\ }\href {https://opg.optica.org/abstract.cfm?uri=optica-11-1-10} {\bibfield  {journal} {\bibinfo  {journal} {Optica}\ }\textbf {\bibinfo {volume} {11}},\ \bibinfo {pages} {10--17} (\bibinfo {year} {2024})}\BibitemShut {NoStop}%
\bibitem [{\citenamefont {Wilson}\ \emph {et~al.}(2015)\citenamefont {Wilson}, \citenamefont {Sudhir}, \citenamefont {Piro}, \citenamefont {Schilling}, \citenamefont {Ghadimi},\ and\ \citenamefont {Kippenberg}}]{wilson2015measurement}%
  \BibitemOpen
  \bibfield  {author} {\bibinfo {author} {\bibfnamefont {D.~J.}\ \bibnamefont {Wilson}}, \bibinfo {author} {\bibfnamefont {V.}~\bibnamefont {Sudhir}}, \bibinfo {author} {\bibfnamefont {N.}~\bibnamefont {Piro}}, \bibinfo {author} {\bibfnamefont {R.}~\bibnamefont {Schilling}}, \bibinfo {author} {\bibfnamefont {A.}~\bibnamefont {Ghadimi}}, \ and\ \bibinfo {author} {\bibfnamefont {T.~J.}\ \bibnamefont {Kippenberg}},\ }\bibfield  {title} {\enquote {\bibinfo {title} {Measurement-based control of a mechanical oscillator at its thermal decoherence rate},}\ }\href {https://doi.org/10.1038/nature14672} {\bibfield  {journal} {\bibinfo  {journal} {Nature}\ }\textbf {\bibinfo {volume} {524}},\ \bibinfo {pages} {325--329} (\bibinfo {year} {2015})}\BibitemShut {NoStop}%
\bibitem [{\citenamefont {Condos}\ \emph {et~al.}(2024)\citenamefont {Condos}, \citenamefont {Pratt}, \citenamefont {Manley}, \citenamefont {Agrawal}, \citenamefont {Schlamminger}, \citenamefont {Pluchar},\ and\ \citenamefont {Wilson}}]{condos2024ultralow}%
  \BibitemOpen
  \bibfield  {author} {\bibinfo {author} {\bibfnamefont {C.}~\bibnamefont {Condos}}, \bibinfo {author} {\bibfnamefont {J.}~\bibnamefont {Pratt}}, \bibinfo {author} {\bibfnamefont {J.}~\bibnamefont {Manley}}, \bibinfo {author} {\bibfnamefont {A.}~\bibnamefont {Agrawal}}, \bibinfo {author} {\bibfnamefont {S.}~\bibnamefont {Schlamminger}}, \bibinfo {author} {\bibfnamefont {C.}~\bibnamefont {Pluchar}}, \ and\ \bibinfo {author} {\bibfnamefont {D.}~\bibnamefont {Wilson}},\ }\bibfield  {title} {\enquote {\bibinfo {title} {Ultralow loss torsion micropendula for chipscale gravimetry},}\ }\href {https://doi.org/10.48550/arXiv.2411.04113} {\bibfield  {journal} {\bibinfo  {journal} {arXiv preprint arXiv:2411.04113}\ } (\bibinfo {year} {2024})}\BibitemShut {NoStop}%
\bibitem [{\citenamefont {Kryhin}\ and\ \citenamefont {Sudhir}(2025)}]{kryhin2023distinguishable}%
  \BibitemOpen
  \bibfield  {author} {\bibinfo {author} {\bibfnamefont {S.}~\bibnamefont {Kryhin}}\ and\ \bibinfo {author} {\bibfnamefont {V.}~\bibnamefont {Sudhir}},\ }\bibfield  {title} {\enquote {\bibinfo {title} {Distinguishable consequence of classical gravity on quantum matter},}\ }\href {https://journals.aps.org/prl/abstract/10.1103/PhysRevLett.134.061501} {\bibfield  {journal} {\bibinfo  {journal} {Phys. Rev. Lett.}\ }\textbf {\bibinfo {volume} {134}},\ \bibinfo {pages} {061501} (\bibinfo {year} {2025})}\BibitemShut {NoStop}%
\end{thebibliography}%
  \end{document}